# Stochastic Approach to Determining the Mass Standard Based on the Fixed Values of Fundamental Physical Constants

Mikhail Batanov-Gaukhman

Ph.D., Associate Professor, Institute No. 2 "Aircraft, rocket engines and power plants",
Federal State Budgetary Educational Institution of Higher Education "Moscow Aviation Institute
(National Research University)", Volokolamsk highway 4, Moscow, Russian Federation
(alsignat@yandex.ru)

**Abstract:** It is shown that the inert properties of a stationary random process can be expressed in terms of the ratio of its correlation interval $\tau_x$ to the doubled variance $D_x$. When using a fixed value of Planck's constant $h$ as a proportionality factor, the ratio $h\tau_x/2D_x$ has the dimension of a kilogram and can be used as an equivalent of a mass standard. It is proposed to use thermal (i.e. Johnson–Nyquist) noise as a reference Gaussian stationary random process. The theoretical justification of the project of creating "Thermal superconducting ampere balance" for measuring the energy mass $m_E$ of an object is also given

## 1 INTRODUCTION

The decisions of the 26th General Conference on Weights and Measures (GCWM), dedicated to the "Revision of the International System of Units (SI)", summed up the implementation of the program for "immaterial" definitions of standards of basic physical quantities.

Starting from May 20, 2019, to implement the new definition of units of measurement, it is necessary to use any equations of physics that relate the mass, Planck's constant $h = 6.62607015 \cdot 10^{-34}$ Js, the speed of light in vacuum $c = 299792458$ m/s, the frequency of the hyperfine transition of the ground state of the cesium-133 atom $v_{Cs} = 9192631770$ Hz, the elementary charge $e = 1.602176634 \cdot 10^{-19}$ C, the Boltzmann constant $k_B = 1.380\,649 \cdot 10^{-23}$ J/K and the Avogadro constant $N_A = 6.02214076 \cdot 10^{23}$ mol$^{-1}$.

One of the main directions in the implementation of the mass standard based on the new definition of SI is the development of two projects "Kibble balance" and "Counting atoms in a quasi-ideal ball of silicon-28 ($^{28}$Si)" (hereinafter "Counting atoms") [4,5]. This area of research is also called «The X-ray-crystal-density (XRCD) method».

"Kibble balance" is based on the relationship between gravitational mass and Planck's constant [1,2]

$$m_g = \frac{jd^2 f^2 h}{4gv},  \qquad (1)$$

where

$j$ is an integer associated with the Landau factor in the quantum Hall effect;

$d$ is an integer associated with the Shapiro steps in the Josephson effect;

$f$ is the excitation frequency of the transition between two superconductors in the Josephson effect.

$v$ is the speed of movement of the winding in a constant magnetic field.

$g$ is the free fall acceleration at the location of the watt balance.

"Counting atoms" (or XRCD method) is due to the proportionality between the inert mass of $^{28}$Si and the Avogadro constant [4]

$$m_i = m_{Si} v_{Si} N_A, \qquad (2)$$

where $m_{Si}$ is the atomic mass of $^{28}$Si; $v_{Si}$ is the number of moles of $^{28}$Si in 1 kg.

These projects are related to each other, because without quasi-ideal $^{28}$Si balls having a mass of 1 kg with a relative error of about $10^{-8}$, it is impossible to compare the readings of "Kibble balances" located in various laboratories around the world. On the other hand, without the "Kibble balance" it is impossible to verify the equivalence of the inert mass of a quasi-ideal $^{28}$Si ball to its gravitational mass [4].

In other words, the weight of each quasi-ideal $^{28}$Si ball must be balanced by Kibble balances with a maximum allowable uncertainty of no more than $10^{-8}$ kg. In this case, quasi-ideal $^{28}$Si balls, which should be included in the Kibble balance, can be used to compare the readings of similar watt-balances located in different places on the planet.

The presence of only one Kibble scale on the planet will lead to similar problems with the international prototype of the kilogram held by the International Bureau of Weights and Measures.

Both of the above projects are an undoubted progress in the field of practical metrology, because their accuracy increases by an order of magnitude the accuracy of the international prototype of the kilogram, stored in the International Bureau of Weights and Measures. Also, these projects are reproducible in any laboratory in the world based on the above fundamental physical principles and fixing the numerical values of the constants $h$ and $N_A$.

At the same time, these projects have a number of shortcomings [4].

The Kibble balance is one of the most complex measuring systems in the world, because in addition to using the superconducting Josephson effect and the quantum Hall effect, it is necessary to measure with high accuracy the speed $v$ of the coil in a uniform magnetic field, as well as the free fall acceleration $g$ at the location of the watt-balance.

The disadvantages of the silicon ball include the possibility of contamination of its surface. In addition, $^{28}$Si crystals are never perfect and monoisotopic [4]. Therefore, it is necessary to take into account corrections for the content of impurities, for defects in the $^{28}$Si crystal lattice (vacancies and interstices), for the formation of an oxide film, and for adsorbed water molecules by the surface layer of the silicon ball. It is also necessary to take into account the mass defect related to the binding energy of the atoms of the $^{28}$Si single crystal [4].

This article proposes a different, stochastic approach to determining the mass standard, based on the relationship between the ratio of the main averaged characteristics of a stationary random process and the ratio of Planck's constant to the mass of a particle participating in this process.

________________________________________________________________

## 2. Relationship between mass, Planck's constant and the characteristics of a stationary random process

The article [6] considers a stationary random process associated with the change in the projection of the location of a chaotic wandering particle (such as a Brownian particle) with time $t$ (see Figure 1)

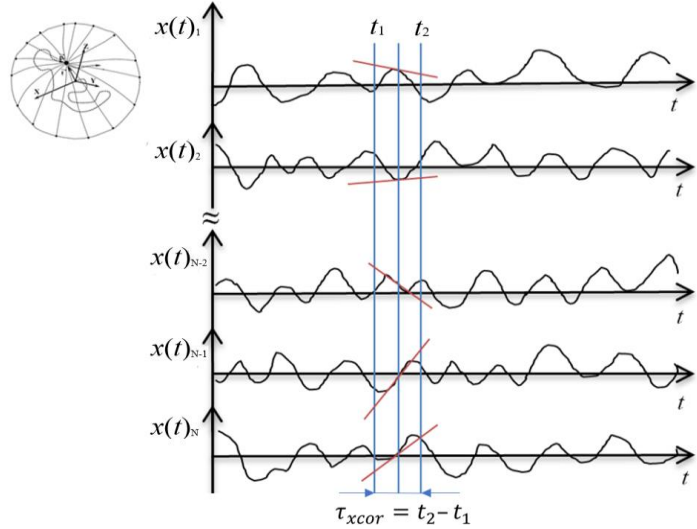

**Fig. 1** $N$ realizations of a stationary random process with an autocorrelation interval $\tau_x \approx \tau_{xcor}$. These implementations can be interpreted, for example, as changes over time $t$ of the projection onto the $X$ axis of the location of a particle randomly wandering in a closed region of 3-dimensional space

As a result of the analysis of this stationary random process $x(t)$ in [6], the following procedure was developed for obtaining the probability density function (PDF) $\rho(x')$ of the derivative of this process with a known one-dimensional PDF $\rho(x)$ of the process itself:

*a*) The initial one-dimensional PDF $\rho(x)$ of a stationary random process is represented as a product of two probability amplitude (PA)

$$\rho(x) = \psi(x)\psi(x) = \psi(x)^2; \qquad (3)$$

*b*) Two Fourier transforms are performed

$$\phi(x') = \frac{1}{\sqrt{2\pi\eta_x}} \int_{-\infty}^{\infty} \psi(x)\, exp\{ ix'x/\eta_x\} dx, \qquad (4)$$

$$\phi^*(x') = \frac{1}{\sqrt{2\pi\eta_x}} \int_{-\infty}^{\infty} \psi(x)\, exp\{ -ix'x/\eta_x\} dx; \qquad (5)$$

*c)* The desired PDF of the derivative of the studied stationary random process is found

$$\rho(x') = \phi(x')\phi^*(x') = |\phi(x')|^2, \tag{6}$$

where, according to Ex. (52) in [6]

$$\eta_x = \frac{2\sigma_x^2}{\tau_x} \tag{7}$$

$\sigma_x$ is the standard deviation of the original stationary random process $x(t)$;

$\tau_x$ is autocorrelation interval of the same random process (see Figure 1).

For example, consider a Gaussian stationary random process $x(t)$. In each section $t_i$ of this process, the random variable $x$ is distributed according to the Gauss law:

$$\rho(x) = \frac{1}{\sqrt{2\pi\sigma_x^2}} exp\{-(x-a_x)^2/2\sigma_x^2\}, \tag{8}$$

where $\sigma_x^2 = D_x$ and $a_x$ are the variance and expected value, resp., of the random process $x(t)$ under study.

According to Ex. (3), we represent PDF (8) as a product of two probability amplitudes

$$\rho(x) = \psi(x)\psi(x),$$

where
$$\psi(x) = \frac{1}{\sqrt[4]{2\pi\sigma_x^2}} exp\{-(x-a_x)^2/4\sigma_x^2\}. \tag{9}$$

Let's substitute PA (9) into Exs. (4) and (5)

$$\phi(x') = \frac{1}{\sqrt{2\pi}} \int_{-\infty}^{\infty} \frac{1}{\sqrt[4]{2\pi\sigma_x^2}} exp\{-(x-a_x)^2/4\sigma_x^2\} exp\{ ix'x/\eta_x\} dx, \tag{10}$$

$$\phi^*(x') = \frac{1}{\sqrt{2\pi}} \int_{-\infty}^{\infty} \frac{1}{\sqrt[4]{2\pi\sigma_x^2}} exp\{-(x-a_x)^2/4\sigma_x^2\} exp\{-ix'x/\eta_x\} dx \tag{11}$$

and perform the integration

$$\phi(x') = \frac{1}{\sqrt[4]{2\pi\eta_x^2/(2\sigma_x)^2}} exp\{-x'^2/[2\eta_x/(2\sigma_x)]^2\} exp\{ ia_xx'/\eta_x\}, \tag{12}$$

$$\phi^*(x') = \frac{1}{\sqrt[4]{2\pi\eta_x^2/(2\sigma_x)^2}} exp\{-x'^2/[2\eta_x/(2\sigma_x)]^2\} exp\{-ia_xx'/\eta_x\}. \tag{13}$$

___

In accordance with Ex. (6), we multiply the PDF (12) and (13), as a result we get

$$\rho(x') = \frac{1}{\sqrt{2\pi\left(\frac{\eta_x}{2\sigma_x}\right)^2}} exp\left\{-\frac{x'^2}{2\left(\frac{\eta_x}{2\sigma_x}\right)^2}\right\} \qquad (14)$$

is PDF of the derivative of the stationary Gaussian random process $x(t)$.

In quantum mechanics, for the transition from the coordinate representation of the position of the particle

$$\rho(x) = \psi(x)\psi^*(x) \qquad (15)$$

a similar procedure is applied to its momentum representation

$$\phi(p_x) = \frac{1}{\sqrt{2\pi\hbar}}\int_{-\infty}^{\infty}\psi(x)\,exp\{ip_x x/\hbar\}dx, \qquad (16)$$

$$\phi^*(p_x) = \frac{1}{\sqrt{2\pi\hbar}}\int_{-\infty}^{\infty}\psi^*\,exp\{-ip_x x/\hbar\}dx, \qquad (17)$$

where in the non-relativistic case (i.e., at low particle speeds compared to the speed of light)

$$p_x = mv_x = m\frac{dx}{dt} = mx', \qquad (18)$$

is the $x$-component of the particle's momentum, which is related to its velocity $v_x$ (i.e., the derivative of its coordinate x with respect to time);

$$\hbar = h/2\pi \qquad (19)$$

is the reduced Planck constant.

The PDF of the $x$-component of the particle momentum is equal to

$$\rho(p_x) = \phi(p_x)\phi^*(p_x) = |\phi(p_x)|^2. \qquad (20)$$

If the position of the particle is described by the Gaussian PDF (8), then performing actions (15) – (20) we obtain

$$\rho(p_x) = |\psi(p_x)|^2 = \frac{1}{\sqrt{2\pi\left(\frac{\hbar}{2\sigma_x}\right)^2}} exp\left\{-\frac{p_x^2}{2\left(\frac{\hbar}{2\sigma_x}\right)^2}\right\}. \qquad (21)$$

Taking into account Ex. (18), we have the PDF of the derivative of the random process under study

$$\rho(x') = \frac{1}{\sqrt{2\pi\left(\frac{\hbar}{m 2\sigma_x}\right)^2}} exp\left\{-\frac{x'^2}{2\left(\frac{\hbar}{m 2\sigma_x}\right)^2}\right\}. \tag{22}$$

Comparing the PDFs of the derivatives (14) and (22), we find that if the stochastic and quantum mechanical approaches considered above refer to the same stationary random process, then

$$\eta_x := \frac{\hbar}{m}. \tag{23}$$

Let's write this relation taking into account Ex. (7)

$$\frac{2\sigma_x^2}{\tau_x} := \frac{\hbar}{m} \quad \left[\frac{m^2}{s}\right]. \tag{24}$$

Both sides of this ratio have the dimension of kinematic viscosity [m²/s].

Relationship (24) implies a connection between the inertial mass of a chaotically wandering particle, the reduced Planck constant and the main characteristics of a stationary random process in which this particle participates

$$m := \frac{\hbar \tau_x}{2\sigma_x^2}. \tag{25}$$

It can be seen from Ex. (25) that the ratio $2\tau_x/\sigma_x^2$ determines the inert properties of a chaotically wandering particle. The larger the autocorrelation interval $\tau_x$ of the random process under study, the smoother the change in the direction of particle motion (i.e., the more inert). On the other hand, the greater the dispersion $D_x = \sigma_x^2$ (characterizing the average deviation of a wandering particle from the average value), the less its resistance (inertia).

Thus, the inert properties of chaotically wandering particle are directly proportional to $\tau_x$ and inversely proportional to $2\sigma_x^2$.

### 3. Statistical approach to the determination of the mass standard

Let's rewrite relation (25) taking into account (19)

$$m := \frac{h \tau_x}{4\pi \sigma_x^2}. \tag{26}$$

The use of relation (26) to determine the mass standard is fully consistent with the resolutions of the 26th GCWM. Indeed, in this expression there is a direct relationship between the mass and Planck's constant $h$.

Therefore, if in nature there is a sufficiently stable stationary random process $x(t)$ with PDF close to Gaussian, then it can be taken as a standard and put it as the basis for determining the standard of inertial mass.

To do this, it is necessary to make two representative samples from this process: $x_{k1}$ (into section $t_1$) and $x_{k2}$ (into section $t_2$) (see Figure 1), at a distance between sections $\tau_{xcor} = t_2 - t_1$ (where $\tau_{xcor}$ is an estimate of the autocorrelation interval $\tau_x$), at which Pearson's autocorrelation coefficient $r(x_{k1}, x_{k2})$ is less than some critical parameter $\varepsilon_{cr}$ steadily tending to zero ($\varepsilon_{cr} \to 0$) (see Figure 2)

$$r(x_1, x_2) = \frac{\sum_{k=1}^{N}(x_{k1}-\overline{x_{k1}})(x_{k2}-\overline{x_{k2}})}{(N-1)\sqrt{D_{x1}D_{x2}}} < \varepsilon_{cr} \to 0, \qquad (27)$$

где  $\overline{x_1}, D_{x1}$  – expected value and variance in cross section $t_1$;

  $\overline{x_2}, D_{x2}$  – expected value and variance in cross section $t_2$.

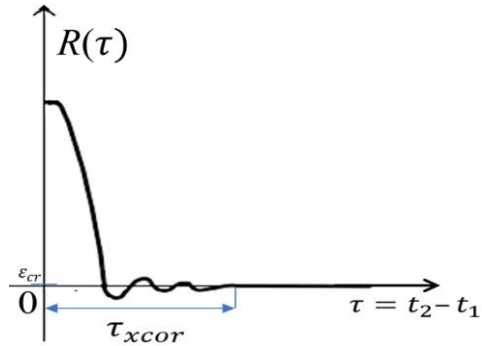

**Fig. 2** The autocorrelation interval $\tau_x \approx \tau_{xcor}$ is the time interval between the values of the correlation function $R(\tau=0)$ and $R(\tau_{xcor}) < \varepsilon_{cr}$, where $\varepsilon_{cr}$ is a small critical parameter steadily tending to zero ($\varepsilon_{cr} \to 0$)

For a stationary random process, the following conditions are met

$$\overline{x_1} = \overline{x_2} = \overline{x}, \quad D_{x1} = D_{x2} = D_x. \qquad (28)$$

Therefore, Ex. (26) can be represented as an approximate equality

$$m \approx \frac{h\tau_{xcor}}{4\pi D_x}. \tag{29}$$

To use Ex. (29) as an equivalent of the mass standard, it is necessary:

1] Choose a stable stationary random process with PDF close to a Gaussian distribution, which can be carried out in any metrological laboratory;

2] Estimate its variance $D_x$ (with a confidence level corresponding to $n\sigma_x = n\sqrt{D_x}$, where $n$ is any natural number providing a given level of accuracy);

3] Estimate the correlation interval $\tau_{xcor}$ at a fixed value of the critical parameter $\varepsilon_{cr}$.

## 3. Reference thermal noise

The thermal noise of the resistor c (i.e., the Johnson-Nyquist noise) is proposed as a reference stationary random process with an almost Gaussian PDF.

The spectral density of the squared average electromotive force (EMF) of thermal noise has the form [8]

$$\langle e_T^2 \rangle_f = 4k_B T R \frac{hf}{k_B T} \left( \exp \frac{hf}{k_B T} - 1 \right)^{-1} \text{ [V}^2\text{/Hz]}, \tag{30}$$

where $T$ is the temperature, $R$ is the resistance.

The graph of this function is shown in Figure 3.

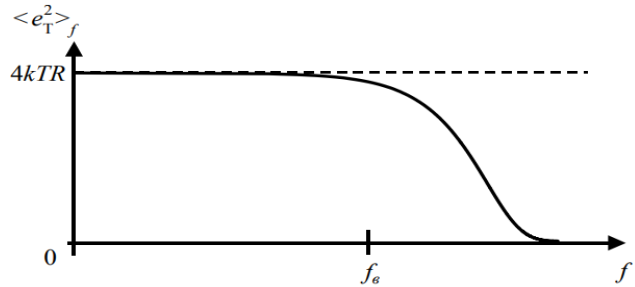

**Fig. 3** The spectral density of squared average EMF of the thermal noise

The autocorrelation function of thermal noise can be obtained using the inverse Fourier transform

$$R(\tau) = \frac{1}{2\pi}\int_{-\infty}^{\infty}\langle e_T^2\rangle_f e^{i2\pi f\tau}df = \frac{1}{2\pi}\int_{-\infty}^{\infty}\frac{4Rh2\pi f}{\left(\exp\frac{h2\pi f}{k_B T}-1\right)}e^{i2\pi f\tau}df.$$

However, in the frequency range for which the inequality

$$\langle e_T^2\rangle_f = 4k_B TR \qquad (31)$$

spectral density (32) can be considered constant, i.e. frequency independent (Nyquist formula) [8]:

$$\langle e_T^2\rangle_f = 4k_B TR. \qquad (32)$$

Therefore, thermal noise in a wide range from 0 Hz to the upper frequency of the $f_в \approx \frac{k_B T}{h}$ [Hz] (for example, at T= 300 K we have $f_в \approx 6\cdot 10^{12}$ Hz) (33) can be viewed as white noise with an autocorrelation function:

$$R(\tau) = \sigma^2\delta(\tau) = 4k_B TR\delta(\tau), \qquad (34)$$

where $\delta\tau$ is the delta function;

$$\sigma^2 = D(e_T) = 4k_B TR \quad [\text{V}^2\text{s}] = \left[\frac{J^2 s}{C^2}\right] \qquad (35)$$

is the variance of the considered stationary random process (i.e. thermal noise).

We multiply Ex. (35) by a constant value, composed of the fundamental physical constants

$$\frac{e^2}{h^2 c^2} \quad \text{with the dimension} \left[\frac{C^2}{m^2 J^2}\right]. \qquad (36)$$

As a result, we get the value

$$\frac{4k_B TRe^2}{h^2 c^2} \quad \text{with the dimension} \left[\frac{s}{m^2}\right]. \qquad (37)$$

Its reciprocal

$$\frac{h^2 c^2}{4k_B TRe^2} \quad \text{has the dimension of kinematic viscosity} \left[\frac{m^2}{s}\right]. \qquad (38)$$

Therefore, the value (38) can be put in accordance with relation (24)

$$\frac{h^2 c^2}{4k_B TRe^2} := \frac{h}{2\pi m} := \frac{2\sigma_x^2}{\tau_x} \left[\frac{m^2}{s}\right]. \qquad (39)$$

This relation allows expressing the mass in terms of thermal noise parameters (i.e. a stable, stationary random process)

$$m := \frac{2k_B T R e^2}{\pi h c^2}. \tag{40}$$

For the most accurate determination of the resistance R, we can use the quantum Hall effect

$$R = \frac{h}{je^2}, \tag{41}$$

where $j$ is an integer associated with the Landau filling factor.

Substituting Ex. (41) into relation (40), we obtain

$$m := \frac{2}{j\pi} \frac{k_B T}{c^2}. \tag{42}$$

This relation corresponds to the formula for the equivalence of the energy $E$ and the mass of A. Einstein

$$m = \frac{E}{c^2}. \tag{43}$$

At the same time, the equivalent of mass $m = 1$ kg corresponds to a large value of thermal energy

$$k_B T = \frac{j\pi}{2} c^2 \approx 14 \cdot 10^{16} \, [J]. \tag{44}$$

However, the International system of units (SI) allows the use of fractional quantities. Therefore, $10^{-16}$ fraction of 1 kg can be taken as the equivalent of the mass standard.

Formally, with fixed values of the fundamental constants $k_B$ and $c$, nothing prevents us from taking as a mass standard the value corresponding to the temperature, for example, the triple point of water $T_i = 273.16$ K, or any other the referent thermal scale points, for example, the crystallization point of aluminum $T_i = 933.473$ K or the crystallization point of copper $T_i = 1357.77$ K. Then expression (42) for $j = 1$ takes the form

$$m_i := \frac{\pi}{2} \frac{k_B T_i}{c^2}. \tag{45}$$

It remains to solve the problem of transferring this unit to measuring instruments.

## 4. Thermoelectric semiconductor ampere balance

For the above task, it is proposed to use an ampere-balance, in which the counterweight to the gravitational mass is created by a heated looped conductor located in a constant magnetic field.

In such a looped conductor, there should be thermal noise with a spectral current density [8]

$$\langle i_T^2 \rangle_f = \frac{4k_B T_i}{R_k} \left[\frac{A^2}{Hz}\right], \qquad (46)$$

where $T_i$ is one of the reference points of the temperature scale.

$R_k = L\rho$ is the resistance of the looped conductor ($L$ is the length of the conductor, i.e. the circumference; $\rho$ is the linear resistance of the conductor).

However, thermal noise in a conductor is a Gaussian multidirectional random process (Figure 4).

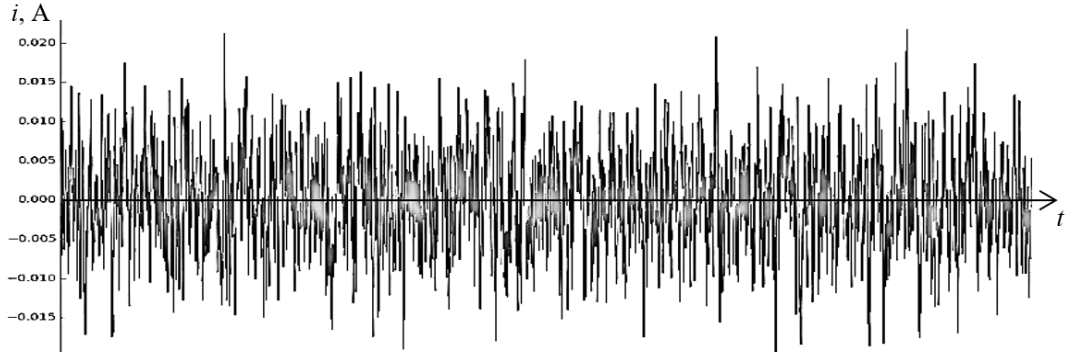

**Fig. 4** Thermal noise in a looped conductor with resistance $R_k$

Therefore, it should be expected that in a constant magnetic field such a looped conductor, on average, will remain motionless.

In this regard, it is proposed to use a semiconductor ring with an equivalent circuit shown in Figure 5.

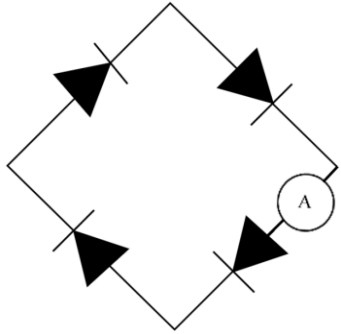

**Fig. 5** The semiconductor ring equivalent circuit

The idea of the semiconductor ring shown in Figure 5, belongs to the MAI master Alexander Bindiman.

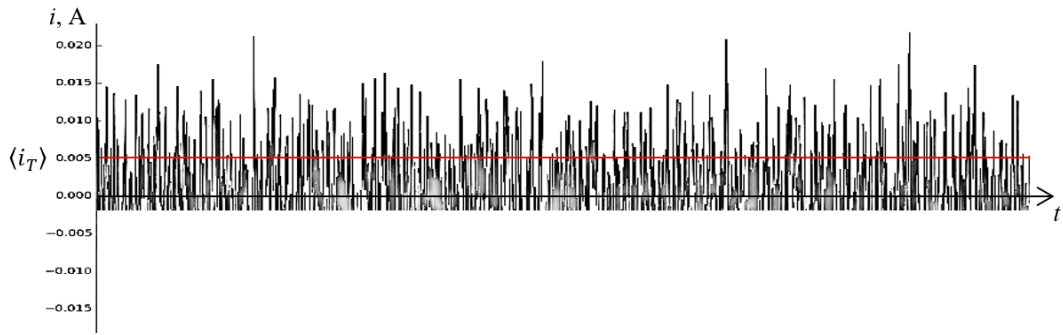

**Fig. 6** Expected thermal noise in a semiconductor ring

In this case, it can be assumed that in such a looped semiconductor, an average direct current should circulate (see Figure 6), approximately equal to the standard deviation of thermal noise

$$\langle i_T \rangle \approx \sqrt{\frac{4k_B T_i}{R_k}} \ [A]. \qquad (47)$$

On closer examination, it should be taken into account that there are potential barriers (i.e., p-n junctions) in a semiconductor ring, therefore, a shot noise component can also exist in it, and other quantum effects are also possible. If such quantum effects are discovered, then the measurement accuracy will increase significantly.

___

The thermoelectric effect can be enhanced by stacking $n$ semiconductor rings isolated from each other. In this case, the total average noise current of a stack of semiconductor rings can be estimated by the expression

$$\langle I_s \rangle \approx n \sqrt{\frac{4k_B T_i}{R_k}} \quad [A]. \tag{48}$$

It is possible that in thin semiconductor rings (several Angstroms thick) located in a constant magnetic field, high-temperature quantum effects, such as the quantum Hall effect, can be observed. If such an effect is detected, then $R_k$ will be determined by an expression like (41), while the accuracy of determining the thermal current (48) will increase significantly

$$\langle I_s \rangle \approx n \sqrt{\frac{4k_B T_i j e^2}{h}} \quad [A]. \tag{49}$$

If this hypothesis is experimentally confirmed, then this thermoelectric effect can be used to create semiconductor ampere scales designed to transfer a unit of weight to a reference (or model) measuring instrument. The scheme of such semiconductor thermoelectric ampere scales is shown in Figure 7.

In such scales, the gravity force $F_1 = mg$ acting on an exemplary load (in this case, a quasi-ideal silicon ball) is compensated by the Ampere force

$$F_2 = k_w \langle I_s \rangle \tag{50}$$

where $k_w$ is the coefficient of proportionality, which functionally depends on the magnetic induction $B$ created by the constant ring magnet (Figure 7) and the length of the active part of the semiconducting ring $l$.

$$k_w \approx Bl$$

The equality of forces $F_1$ and $F_2$, taking into account expression (49), leads to the relation

$$m := \frac{B \langle I_s \rangle l}{g} = \frac{Bln}{g} \sqrt{\frac{4k_B e^2}{h} T_i} \,. \tag{51}$$

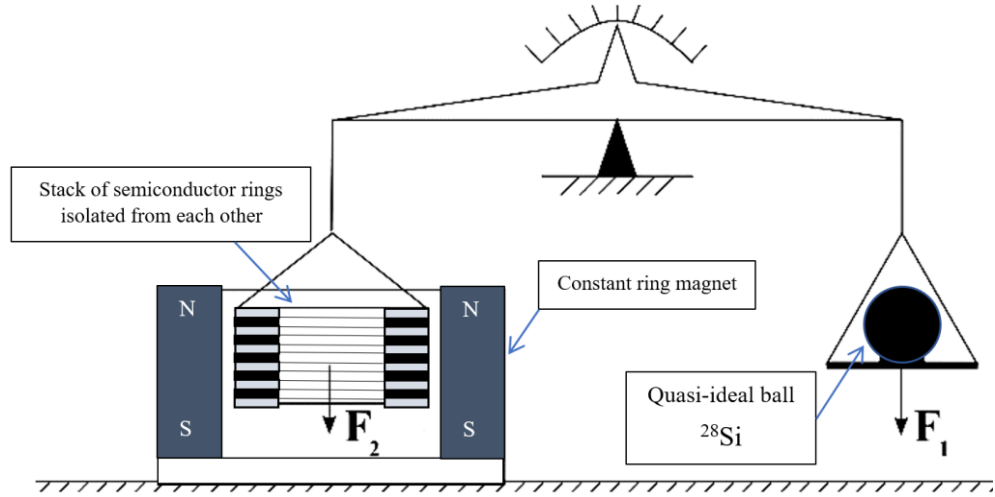

**Fig. 7** The scheme of thermoelectric semiconductor ampere-balances for transferring a measure of mass to a reference or exemplary measuring instrument (in particular, a quasi-ideal silicon ball)

As a reference (or exemplary) measuring instrument, a quasi-ideal $^{28}$Si ball with a mass of $1g$ can be used, with the help of which a unit of mass can be transferred to other measuring instruments, according to an appropriate verification scheme, and comparison of readings with thermoelectric semiconductor amperes - scales located in various laboratories of the world.

Based on expression (51), we can take as a mass standard the value corresponding to the temperature, for example, of zinc crystallization $T_i = 692.677$ K

$$m := \frac{Bln}{g}\sqrt{\frac{4k_B e^2}{h} 692.677}. \qquad (52)$$

Let $B = 1$T, $l = 1$m, $g = 9.81$ m/s$^2$, $n = 1$ then

$$m := \frac{Bl}{g}\sqrt{\frac{4k_B e^2}{h} 692.677} \approx 13 \cdot 10^{-14} \text{ kg} \approx 1.3 \cdot 10^{-10} \text{ g}. \qquad (53)$$

In order for the right side of Ex. (52) to be equal to 1g, it is necessary that the stack of isolated looped semiconductors consists of $n \approx 7 \cdot 10^9$ layers. Indeed, substituting the given value of $n$ into Ex. (52)

$$m := \frac{1 \times 1 \times 7 \cdot 10^9}{9.81}\sqrt{\frac{4k_B e^2}{h} 692.677} \approx 1\text{g}. \qquad (54)$$

If it is possible to obtain a thickness of one insulated semiconductor layer of the order of 1.4Å = 1.4·10$^{-10}$ m, then a stack of $n = 7 \cdot 10^9$ of such layers will turn out to be a height of the order of $H \approx 0.14 \cdot 10^{-9} \times 7 \cdot 10^9 \approx 1\,m$.

The creation of such a stack of isolated looped semiconductors is at the edge of the possibilities of modern nanotechnology. It is possible to increase the thickness of the semiconductor layer with a significant increase in $B$, $T_i$, and a decrease in $R_k$. For example, at $R_k = 100\,\Omega$, $B = 1$T and $T_i = 692.677$ K, we obtain the value

$$m := \frac{Bl}{g}\sqrt{\frac{4T_i k_B}{R_k}} \approx \frac{1}{9{,}81}\sqrt{\frac{4 \times 692{,}677 \times 1{,}380\,649 \cdot 10^{-23}}{100}} \approx 20.6 \cdot 10^{-12}\,\text{kg} \approx 2 \cdot 10^{-8}\,\text{g},$$

(55)

which is more than two orders of magnitude larger than (53). Therefore, the thickness of the looped semiconductor layer in a stack 1m high can be increased by approximately two orders of magnitude up to ~10$^{-8}$ m.

Unfortunately, at this stage of the study, it is impossible to establish the main limitations of the proposed method. The main metrological characteristics of the thermoelectric semiconductor ampere-balances can be established only during the experimental implementation of this project.

**4. Conclusions**

Fixing the numerical values of the fundamental physical constants $h$, $e$, $k_B$, $c$, $N_A$ on the basis of the resolutions of the 26th GCWM opens up wide possibilities for applying various physical principles to determine the standards of physical quantities.

In particular, this article proposes to use the possibility of expressing the ratio of the Planck constant to the mass of a particle through the ratio of the main averaged characteristics of a Gaussian stationary random process: the doubled variance to the autocorrelation coefficient of this process (24). Whence follows the desired dependence (31)

$$m \approx \frac{h\tau_{xcor}}{4\pi D_x}.$$

As a realization of this idea, the theoretical possibility of using thermal noise as a Gaussian stationary random process is considered.

As a result, it is shown that, based on Ex. (51), it is possible to take as a mass standard the value corresponding to the thermodynamic temperature, for example, the triple point of water $T_i = 273.16$ K (or any other reference point on the temperature scale, for example, aluminum crystallization point $T_i = 933.473$ K, or zinc crystallization point $T_i = 692.677$ K)

$$m := \frac{Bln}{g} \sqrt{\frac{4k_B j e^2}{h} T_i} \ . \tag{56}$$

This effect is barely perceptible. For example, at $T_i = 692.677$ K, $B = 1$T, $l = 1$m, g $= 9.81$ m/s², $n = 1$, $j = 1$, according to (53) $m \approx 1.3 \cdot 10^{-10}$g. However, it all depends on what value to take as the mass standard. For example, if we take the Planck mass $M_p = 2.176\ 434 \cdot 10^{-5}$g as a standard, then by selecting the parameters $T_i$, $B$, $l$, $n$, $j$, you can achieve $m \approx M_p$ with a given accuracy at quite acceptable: temperature, magnetic field and thickness of the looped semiconductor layer of the stack (Figure 7).

It is possible that the creation of thermoelectric semiconductor ampere-balances will turn out to be no less complex and expensive project than the Kibble balance. However, the hope for success is inspired by the fact that in formula (56) there are fewer parameters to be measured and controlled than in formula (1). In addition, the measurement of magnetic thrust in the thermoelectric method is carried out in one stage, and in the Kibble balance, the fictitious power is obtained in two stages.

At the same time, the "thermal electric effect" proposed for study is associated with the creation of traction without energy supply from an external power source. In this case, for the occurrence of accelerated motion of a stack of looped semiconductors (Figure 7), it is sufficient to have a constant magnetic field and an environment with a high temperature.

___

The measuring setup shown in Figure 7 involves the development of two projects: 1) the creation of thermoelectric semiconductor ampere-balances; and 2) the creation of a quasi-ideal silicon-28 ball with an inertial mass of about 1g.

Only by combining these two projects into one project will it be possible to provide for verification schemes with an accuracy corresponding to the requirements of the 26th GCWM, and to compare the readings of thermoelectric semiconductor ampere-balances located in various laboratories of the world.

It should be noted that a critical understanding of the concepts of mass and force leads to the conclusion that the concepts of ideal and actual accelerations are more fundamental (Appendix 1). A detailed discussion of this issue is given in [11].

**Acknowledgement**

Appendix 1

**A1 Exclusion of the concept of mass**

Let us return to consideration of Ex. (31)

$$m \approx \frac{h\tau_{xcor}}{4\pi D_x}. \tag{A1}$$

Obviously, the inertia of the stochastic system under study is related to the ratio of the main averaged characteristics of the stationary random process $\tau_{xcor}/D_x$ with the dimension [s/m$^2$].

The concept of mass (*m*) in this case is introduced artificially by means of the dimensional constant *h*.

In other words, to describe the inert properties of a body, it is not necessary to introduce the concept of mass. Often, by the trajectory of a compact object, one can judge its resistance to force influences that affect a change in its state.

Similarly, it can be shown that the gravitational mass is an unnecessary concept, which is introduced using a dimensional constant – the gravitational constant $G$, and the energy mass is artificially introduced using a dimensional constant – the Boltzmann constant $k_B$.

In connection with the foregoing, the question is quite reasonable and appropriate: "Maybe we should completely get rid of the concept of mass with the dimension of a kilogram?"

This will require a complete revision of the system of units of measurement, but at the same time, science will cleared of many superfluous and insufficiently defined concepts such as force, energy, momentum, etc.

The development of the massless concept can begin with the reformulation of the three laws of classical Newton mechanics:

| **Mass-dependent concept** | **Massless concept** |
|---|---|
| *Newton's First Law* $$v_x = \frac{dx}{dt} = const,$$ when $F_x = 0$ | *First Law of the mass independent mechanics* $$v_x = \frac{dx}{dt} = const, \qquad (A2)$$ when $a_x = 0$ |
| *Newton's Second Law* $$F_x = ma_x' \ [N],$$ where $F_x$ is the $x$-component of the force vector; $a_x' = \frac{d^2x}{dt^2}$ is the $x$-component of the "ideal" acceleration | *Second law of the mass independent mechanics* $$a_x = \mu_x a_x' \ \left[\frac{\text{M}}{\text{C}^2}\right], \qquad (A3)$$ where $a_x$ is the $x$-component of the "actual" acceleration, taking into account the resistance to a change in the state of motion; $a_x' = \frac{d^2x}{dt^2}$ is $x$-component of the "ideal" acceleration; $\mu_x$ is a dimensionless resistance coefficient that connects the "actual" and "ideal" accelerations |
| *Newton's Third Law* Action equals reaction $$F_x = -F_x \ [N].$$ | *Third law of the mass independent mechanics* $$a_x = -a_x \ \left[\frac{\text{M}}{\text{C}^2}\right]. \qquad (A4)$$ |

---

Obviously, with a mass-independent formulation, the meaning and properties of mechanics do not change at all. Only the superfluous, unclear concept of "mass" disappears, which, based on the principle of "Occam's razor", should be discarded.

The reason for inertia is connected with the properties of the space around us. Therefore, in order to formulate the massless concept at a more fundamental level, consider the propagation of a light beam in vacuum with a speed $c$, which is described by the square of the interval

$$ds^2 = c^2 dt^2 - dx^2 - dy^2 - dz^2. \tag{A5}$$

Let's take the value of $c^2 dt^2$ out of brackets, and extract the root from the two sides of the resulting expression

$$ds = cdt\sqrt{1 - \frac{v^2}{c^2}}, \tag{A6}$$

where

$$v = (dx^2 + dy^2 + dz^2)^{1/2}/dt = dl/dt \quad \left[\frac{M}{c}\right] \tag{A7}$$

is the speed of the light source relative to vacuum.

The 4-velocity vector in the framework of the special theory of relativity is defined by the expression

$$u_i = dx_i/ds. \tag{A8}$$

Substitute Ex. (A6) into (A8), as a result we obtain

$$u_i = \left[ \frac{1}{\sqrt{1-\frac{v^2}{c^2}}}, \frac{v_x}{c\sqrt{1-\frac{v^2}{c^2}}}, \frac{v_y}{c\sqrt{1-\frac{v^2}{c^2}}}, \frac{v_z}{c\sqrt{1-\frac{v^2}{c^2}}} \right]. \tag{A9}$$

Differentiating the components of the 4-velocity vector (A9) with respect to $ct$, we obtain the components of the 4-acceleration vector

$$\frac{du_i}{cdt} = \left[ \frac{d}{cdt}\left(\frac{1}{\sqrt{1-\frac{v_x^2}{c^2}}}\right), \frac{d}{cdt}\left(\frac{v_x}{c\sqrt{1-\frac{v_x^2}{c^2}}}\right), \frac{d}{cdt}\left(\frac{v_x}{c\sqrt{1-\frac{v_x^2}{c^2}}}\right), \frac{d}{cdt}\left(\frac{v_x}{c\sqrt{1-\frac{v_x^2}{c^2}}}\right) \right]. \tag{A10}$$

Consider the $x$-component of the 4-acceleration vector

$$\frac{du_x}{cdt} = \frac{d}{cdt}\left(\frac{v_x}{c\sqrt{1-\frac{v_x^2}{c^2}}}\right), \tag{A11}$$

where the value

$$\frac{d}{dt}\left(\frac{v_x}{\sqrt{1-\frac{v_x^2}{c^2}}}\right) = a_x \left[\frac{M}{c^2}\right] \tag{A12}$$

has the dimension of the *x*-component of the 3-acceleration vector.

Let's perform the differentiation operation on the left side of the Ex. (A12)

$$a_x = \left(\frac{1}{\sqrt{1-\frac{v_x^2}{c^2}}} + \frac{v_x^2}{c^2\left(1-\frac{v_x^2}{c^2}\right)^{\frac{3}{2}}}\right)\frac{dv_x}{dt}, \tag{A13}$$

and introduce the notation

$$dv_x/dt = a'_x. \tag{A14}$$

In this case, Ex. (A13) takes the form

$$a_x = \left(\frac{1}{\sqrt{1-\frac{v_x^2}{c^2}}} + \frac{v_x^2}{c^2\left(1-\frac{v_x^2}{c^2}\right)^{\frac{3}{2}}}\right)a'_x, \tag{A15}$$

where  $a_x$ is the "actual" acceleration;

$a_x'$ is the "ideal" acceleration.

The resulting Ex. (A15) has the form of the Second law of mass-independent mechanics (A3)

$$a_x = \mu_x a'_x, \tag{A16}$$

where

$$\mu_x = \left(\frac{1}{\sqrt{1-\frac{v_x^2}{c^2}}} + \frac{v_x^2}{c^2\left(1-\frac{v_x^2}{c^2}\right)^{\frac{3}{2}}}\right) \tag{A17}$$

is a dimensionless coefficient of vacuum resistance linking the "actual" and "ideal" accelerations of the light source.

The coefficient $\mu_x$ shows how the vacuum resistance changes when the speed of movement of the light source changes in the direction of the *x* axis.

From the expression (A17) it can be seen that at the very beginning of the movement (i.e. at $v_x = 0$) the resistance is $\mu_x = 1$. Whereas at $v_x = c$, the vacuum becomes an almost infinite obstacle to further increase in the velocity of the source (i.e., $\mu_x = \infty$).

This is similar to the beginning of the movement of a solid body in the air. At the beginning, the air practically does not manifest itself, however, as the speed of the body increases, the air resistance increases and reaches a maximum at a body speed close to the speed of sound.

The coefficient $\mu_x$ shows how the vacuum resistance changes with a change in the speed of the light source in the direction of the *x* axis.

Vacuum resistance (A17) is associated with the limitation of the speed of light, and this property of the space around us underlies all physical phenomena.

*It is interesting to note that in SI the unit of resistance is "Ohm", and in Hinduism the mantra "Om" (ॐ) is the original sacred sound vibration, by which the entire Universe was created.*

Therefore, the consistent construction of the theory (statics, kinematics and dynamics) based on the dimensionless concept of "resistance of the vacuum" will naturally lead to the complete geometrization of all physical views (as shown in [11]), and only the speed of light in vacuum will remain among the dimensional fundamental constants and the parametric radius of the Universe [11].

The exclusion of the outdated heuristic concept of "mass" will lead to significant progress in the philosophical rethinking of the surrounding reality, and a completely geometrized physics cleared of this concept can lead to significant technological breakthroughs in various branches of knowledge.

_________________________________________________________________________________